\pdfoutput=1
\documentclass[journal]{IEEEtran}
\usepackage{amsmath,amsfonts}
\usepackage{algorithm}
\usepackage{array}
\usepackage[caption=false,font=normalsize,labelfont=sf,textfont=sf]{subfig}
\usepackage{textcomp}
\usepackage{stfloats}
\usepackage{url}
\usepackage{verbatim}
\usepackage{graphicx}
\usepackage{cite}
\usepackage{algorithmic}

\usepackage{amssymb,mathtools} 
\usepackage{siunitx} 
\usepackage{stackengine} 
\usepackage{booktabs} 
\usepackage{url}

\newcommand{\floor}[1]{\left\lfloor #1 \right\rfloor} 
\newcommand\oast{\stackMath\mathbin{\stackinset{c}{0ex}{c}{0ex}{\ast}{\bigcirc}}}

\begin{document}

\title{Generative Adversarial Networks for Pseudo-Radio-Signal Synthesis}

\author{
Haythem Chaker,~\IEEEmembership{Graduate Student Member,~IEEE,}
Soumaya Hamouda,~\IEEEmembership{Senior Member,~IEEE,}\\
and Nicola Michailow
\thanks{
This work has been submitted to the IEEE for possible publication. Copyright may be transferred without notice, after which this version may no longer be accessible.

H. Chaker is with the SIGCOM research group at the Interdisciplinary Centre for Security, Reliability and Trust (SnT), University of Luxembourg, Luxembourg (\textit{corresponding author: haythem.chaker@uni.lu}). S. Hamouda is with the MEDIATRON research laboratory at the Telecommunications Engineering School of Tunis (Sup’Com), University of Carthage, Tunisia. N. Michailow is with Siemens Technology, Munich, Germany.
}
}



\maketitle

\begin{abstract}
For many wireless communication applications, traffic pattern modeling of radio signals combined with channel effects is much needed. While analytical models are used to capture these phenomena, real world non-linear effects (e.g. device responses, interferences, distortions, noise) and especially the combination of such effects can be difficult to capture by these models. This is simply due to their complexity and degrees of freedom which can be hard to explicitize in compact expressions.
In this paper, we propose a more model-free approach to jointly approximate an end-to-end black-boxed wireless communication scenario using software-defined radio platforms and optimize for an efficient synthesis of subsequently similar “pseudo-radio-signals”. More precisely, we implement a generative adversarial network based solution that automatically learns radio properties from recorded prototypes in specific scenarios. This allows for a high degree of expressive freedom. Numerical results show that the prototypes' traffic patterns jointly with channel effects are learned without the introduction of assumptions about the scenario or the simplification to a parametric model.
\end{abstract}


\section{Introduction}
\IEEEPARstart{R}{adio}
signals are all around us and they serve as a key enabler for both communications and sensing, as our daily, commercial and industrial needs increasingly grow reliant on a heavily interconnected and automated world. Over the past century, much effort has gone into expert system design and optimization for both wireless communication and localization systems \cite{principles2}. The main considerations are on how to precisely represent, shape, adapt, and recover these signals through a lossy, non-linear, and often interference heavy channel environment.

In the recent years on an other hand, heavily expert-tuned basis functions, such as Gabor filters in the vision domain, have been largely discarded due to the speed at which they can be naively learned and adapted using feature learning approaches in deep neural networks \cite{R5}.

In this paper, we aim to characterize wireless traffic patterns. We explore making a similar transition from using expert-designed representations and models of wireless links towards using a data driven approach to generate waveforms related to an arbitrary end-to-end wireless communication traffic pattern.

For instance, in context of product testing, network planning or network optimization, R\&D engineers usually turn to network and waveform simulators (e.g. \cite{gnuradio}) that use tractable channel models and have solid foundations on information theory and statistics. Past experience has shown that, in such an approach, data traffic and waveform generation can prove to be time consuming and not efficient for all scenarios. This is because, in simulators, not all details of wireless standards are implemented. Moreover, once transmitted over-the-air, the signal passes through many harsh effects imbued with the chaotic and random laws of nature.

To expand the scope of wireless signal emulation beyond purely theoretical analysis, real radio signals can be digitally recorded and processed using state-of-the-art software-defined radio (SDR) hardware platforms (e.g. \cite{R1}). The recorded signal includes the end-to-end wireless communication scenario characteristics, namely:
\begin{itemize}
    \item The traffic pattern (e.g. modulation and multiple access schemes),
    \item Setup configuration (e.g. distance between the devices and signal attenuation),
    \item Radio frequency (RF) medium particularities (e.g multipath propagation and thermal noise),
    \item Hardware imperfections (e.g. non-linear amplification and finite resolution sampling).
\end{itemize}


With the goal of retransmission, populating a database of real traffic pattern prototypes recorded with SDR is cost-effective and can replace software-over-the-air simulators. However, recordings of raw wireless signals at sampling rates as high as \SI{160}{MHz}, in real-time, challenge the read/write speeds and storage capacities of modern commercial storage solutions \mbox{(around $\SI{1.28}{GByte}$ for  $\SI{160}{MSamples /s}$)}.

In this work, the storage and timelessness issues are treated as a signal generation problem. The idea is to digitally store a relatively short prototype recording of a real wireless signal under a certain traffic pattern scenario, and further develop a machine learning (ML) method able to synthesize a \emph{pseudo-radio-signal} (with an arbitrary length) that has the same radio properties as the prototype. This would particularly solve the storage problem by building, for different scenarios, a catalogue of lightweight signal generation models that are non-parametric, i.e., with no prior information on the channel.

Nevertheless, channel modeling happens to be critical, not only in the design and evaluation but also when learning \emph{online} over a wireless communication chain.
Related works \cite{R1,R2,R3,R4,R5,R6} regarding the design and learning involving end-to-end wireless systems have focused on using simplified analytics such as the additive white Gaussian noise and the Rayleigh fading channel models. Recent works \cite{GAN7,GAN8,GAN9,GAN10,GAN11} tackled channel agnostic over-the-air learning approaches (with no simulated approximations) to learn modulation schemes from real world measurements using SDR hardware platforms and capabilities.
Historically, the latter method is enabled following the introduction of generative adversarial networks (GAN) \cite{goodfellow-gan}.

GANs have been extensively applied in computer vision and text analytics to generate synthetic data that is statistically similar to real data \cite{goodfellow-dl}. More recently, there have been efforts to apply GANs to wireless communications \emph{offline} too to augment training (waveform) data sets such as those used to train classifiers for spectrum sensing and jamming \cite{GAN10}.

To solve the signal generation problem, we propose a GAN-based method along with its parameter tuning and the digital signal processing (DSP) necessary to efficiently command the pseudo-radio-signal synthesis operation. The geometric disposition of radio devices in a prototyped scenario are known as domain knowledge, while no assumptions are made on the channel where the traffic pattern is occurring (e.g. carrier(s) and bandwidth(s) of interest).

This paper is organized as follows. In Section \ref{sec:system-model-pf}, the system model and problem formulation are introduced. In Section \ref{sec:gan0}, GAN basics and advantages are elaborated. In Section \ref{sec:prss-methodology}, the pseudo-radio-signal synthesis operation is detailed. The structure of the proposed GAN model is explored in Section \ref{sec:proposed}. Numerical results and outlook are presented in Section \ref{sec:numerical-res}. Conclusions are drawn at the end of this paper in Section \ref{sec:conclusion}.

\section{System Model and Problem Formulation}\label{sec:system-model-pf}
\subsection{System model}
Consider a simple device-to-device wireless communication system where a transmitter $T$ interacts with a receiver $R$ over-the-air through a radio channel $H$.

Let the transmitted signal $s(t)$ incorporate detectable wavefroms typical in digital communication and related to modulation, coding and multiple access schemes. The properties of $s(t)$ then denote the traffic pattern particular to the end-to-end wireless communication between $T$ and $R$. Before being captured as $x(t)$, the waveform undergoes joint transformations such as path-loss, multi-path propagation effects, thermal noise as well as other hardware imperfections like phase shifts and non-linear amplifications all depicted in $h(t)$.

For sake of our proposed method, we add to the system an SDR platform able to capture $x(t)$ but with no prior synchronization nor negotiations with the transmitter $T$. Assuming wideband SDR capabilities, the signal $x(t)$ is recorded for a duration $T_\text{prototype}$ and is dubbed the scenario prototype.

The analog signal $x(t)$ captures the joint non-linearities between the stochastic signal $s(t)$ and the unknown channel response $h(t)$ characterizing the black-box end-to-end wireless link between $T$ and the SDR (playing the role of $R$) placed at a fixed seperation distance.
Our objective is to generate a \mbox{pseudo-radio-signal} $x^\prime(t)$ for an arbitrary duration \mbox{$T_\text{gen} > T_\text{prototype}$}, while keeping the same radio properties of the prototype.
In other words, the task is to find $x^\prime(t)$ with a response at $R$ that is statistically similar to the prototype's $x(t)$ (once we broadcast it with SDR). Therefore a correct eventual solution would be to synthesize
\begin{equation} 
    x' \sim p_\text{prototype}(x'). \label{eq:0}
\end{equation}
In traditional engineering workflow, finding the continuous probability distribution $p_\text{prototype}$ is a \emph{model deficit} problem \cite{SUR3} as no physics-based parametric mathematical model of $h(t)$ can be extracted (even in the absence of diagnosable traffic).

On an other hand, for sake of generality, the transmitter device $T$ and its distance from the SDR hardware can differ from scenario to another which entails different prototype measurements for a same traffic pattern. SDR platforms can be purposefully suited for this: where digital spectrum sensing and synchronization efforts can be done to identify the type of the device $T$ and the nature of the detected waveforms for a number of cases and within a range of possible separation distances (see for example \cite{R1}). This, however, is too complex to design for all wireless systems and use-cases which would limit the generalization of the problem at hand. Therefore, \mbox{pseudo-radio-signal} synthesis is also an \emph{algorithm deficit} problem for which lower-complexity solutions are envisaged.

\subsection{Problem formulation}
In these model and algorithm deficit contexts, the relationship \eqref{eq:0} is interpreted as a density approximation problem. In the digital domain, let the stochastic process \mbox{$X=\{x=x\left(\frac{t}{R_\text{s}}\right)\}_{0<t<T_\text{prototype}}$} sampled at a sampling rate $R_\text{s}>0$ yield $N_\text{prototype}$ observations $x$. We formulate the density approximation statistically by constructing an objective model defined by a probability distribution $p_\text{model}$ and parameters $\theta$, to be used to sample $x'$, such that
\begin{equation}
    \begin{aligned}
        X&=\{x \mid x \sim p_\text{prototype}(x)\},~|X|=N_\text{prototype};\\
        &p_\text{model}(x;\theta) \approx p_\text{prototype}(x).  \label{eq:pb}
    \end{aligned}
\end{equation}

\begin{figure*}
	\centering
	\includegraphics[width=\textwidth]{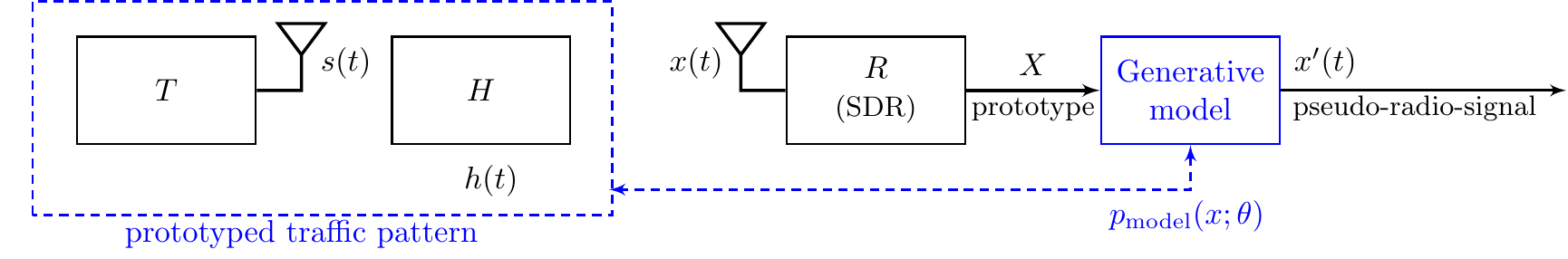}
	\caption{High level formulation of the generative modeling problem for pseudo-radio-signal synthesis.}
	\label{fig:ticv2}
\end{figure*}

Finding the probabilistic model $p_\text{model}$ with optimal parameters $\theta^*$ is called a \emph{generative modeling} problem \cite{generative} as it allows to synthesize $x'$. By resolving this problem, the traffic pattern in $s(t)$ and the natural channel effects $h(t)$ can be substituted by such an optimal waveform sampler $p_\text{model}(\cdot~;\theta^*)$ to synthesize pseudo-radio-signal $x'(t)$ of chosen length $T_\text{gen}>T_\text{prototype}$ using DSP. Fig.~\ref{fig:ticv2} gives a high level formulation of this generative modeling problem for pseudo-radio-signal synthesis. This technique is useful for many radio signal applications including traffic pattern emulation.

If domain knowledge was available on $x(t)$, maximum likelihood estimation (MLE) (or Kullback–Leibler divergence) could be used to explicitly approximate the density $p_\text{model}$. The problem could then be solved using MLE inference equations in an auto-regressive manner. However, explicit density estimation techniques are discarded in our case because as we formulate it, the problem is agnostic to the wideband channel and the transmit-receive chain is treated as a black box. In consequence, no a priori information would be available.

On this account, the density $p_\text{model}$ must be implicitly estimated \cite{generative}. This means that a solution must be found without explicitly defining the density function $p_\text{model}(x;\theta)$. In a probabilistic framework, consider the large set $\mathcal{S}$ of possible samples featuring the traffic pattern “dynamics”. The solution we target would optimally limit the (larger) boundary of space $\mathcal{S}$ to the one of space $X$.

Recent advances in computational resources \cite{SUR3,SUR4} allow to investigate data-driven methods, namely deep neural networks (DNN) mechanisms, to processes raw data, e.g. our prototype $X$ \cite{SUR1}, as input in order to analyse its high-dimensional dynamics. In our case, these dynamics are consist of the captured joint non-linearities between $s(t)$ and $h(t)$ that virtually describe the prototyped traffic pattern. In fact, the minimization of the arbitrary sampling space $\mathcal{S}$ towards the prototyped dynamics space $X$ is the problem we aim to tackle using a deep learning (DL) approach. If the DL model is successfully tested and validated to output samples with magnitudes following the distribution $p_\text{model}$, it is named a \mbox{\emph{deep generative model}}. 

With proper DSP, a deep generative model substitutes the step of acquiring domain knowledge on the prototyped traffic pattern with the potentially easier task of collecting a sufficient number of prototype samples $N_\text{s}$ to conceive a generative algorithm of interest. Learning is made possible by the choice of a set of possible “DNN machines” with a certain process that optimizes their “\text{trainable} parameters” $\theta_\text{DNN}$ in order to build $p_\text{model}(x,\theta^*) \approx p_\text{prototype}(x)$. In the next section, we explore GAN: a (modern) deep generative model that implicitly estimate $p_\text{model}$ directly.

\section{GAN Basics}\label{sec:gan0}
\subsection{Adversarial function approximation}\label{sec:afa}

To implicitly estimate $p_\text{prototype}(x)$, the prototype $X$ is processed, first, by a DNN binary classifier $D(.,\theta_D)$ (with two outputs), henceforth termed the “discriminator”. This DNN is described with the trainable parameters $\theta_D$ and aims to distinguish samples of $X$ from a prior random noise $z \sim p_z(z)$. The noise samples $z$ are the output of a second DNN $G(z,\theta_G)$, henceforth called the “generator”, described with the trainable parameters $\theta_G$.

In a first version, called pre-training, the discriminator \emph{alone} is trained to maximize its ability to distinguish its input between the two categories: i) from $X$ or ii) not from $X$.

The classification results are then sent to the generator as feedback. The latter then manipulates $p_z$ in order to synthesize samples $x'=G(z,\theta_G)$ with similar properties to the input samples $x$ of the prototype $X$. Consequently, this \emph{minimizes} the classification success of the discriminator. This process continues offline (in training) with joint updates on $G(z,\theta_G)$ and $D(.,\theta_D)$.

In game theory, the described design is called a minimax game because it involves minimization in an outer loop and maximization in an inner loop. Accordingly, simultaneous training of the two adversary players is guided by a minimax value function $V$, i.e.:

\begin{equation}
	\begin{aligned}
	(\theta_D^*,\theta_G^*)=\arg \min_{G} \max_{D} V(G,D).  \label{eq:game}
	\end{aligned}
\end{equation}

The game is implemented using an iterative numerical approach with updates on $(\theta_D,\theta_G)$: the trainable parameters of the two DNNs. In theory, if the generator and the discriminator are given enough capacity in the non-parametric limit, a convergence criterion allows to recover the data generating distribution $p_\text{model}(x,\theta^*)$ on the generator's output. An optimum $(\theta_D^*,\theta_G^*)$ is called Nash equilibrium \cite{goodfellow-dl} of the game. At Nash equilibrium, the generator is said to be playing the role of a density estimator of $p_\text{prototype}(x)$ as it would able to output samples identical to the prototyped samples. In this case, the discriminator would distinguishes between the two classes i) and ii) with an equal probability of $\frac{1}{2}$.

In practice, however, $p_\text{prototype}(x)$ and $p_\text{G}(x)$ are not always non-zero, therefore the success ratio of the discriminator is numerically unstable and $D(x)$ itself is typically not convex for all values of $x$. Hence, finding the optimum in $D(x)$ is neither unique nor guaranteed. This is why, in parameter space, adversarial training on $(\theta_D,\theta_G)$ makes the performance of $D(x,\theta_D)$ get closer to $D^*(x,\theta_D^*)$ upon a stop criterion: at which the best associated generator $G^*(x,\theta_G^*)$ implicitly approximates $p_\text{prototype}(x)$.

Optimizing the discriminator to completion in the inner loop in function space is computationally prohibitive \cite{goodfellow-gan}. Moreover, a finite number of inputs would result in overfitting. Since the updates happen on $(\theta_D,\theta_G)$ in the DNN parameters space, and not in the function space, this discards convexity issues but introduces multiple critical points dependent on both DNN architectures. In fact, choosing $p_\text{model}(x,\theta)$ is limited by a family of possible densities $p_\text{G}$ that represent a space $X^\prime=\{x^\prime \sim p_{G(z,\theta_G)}(x^\prime) \mid \forall \theta_G\}$ constructable via any function $G(z;\theta_G)$ instead of $\theta$.

\begin{figure}[t] 
	\centering
	\includegraphics[width=\columnwidth]{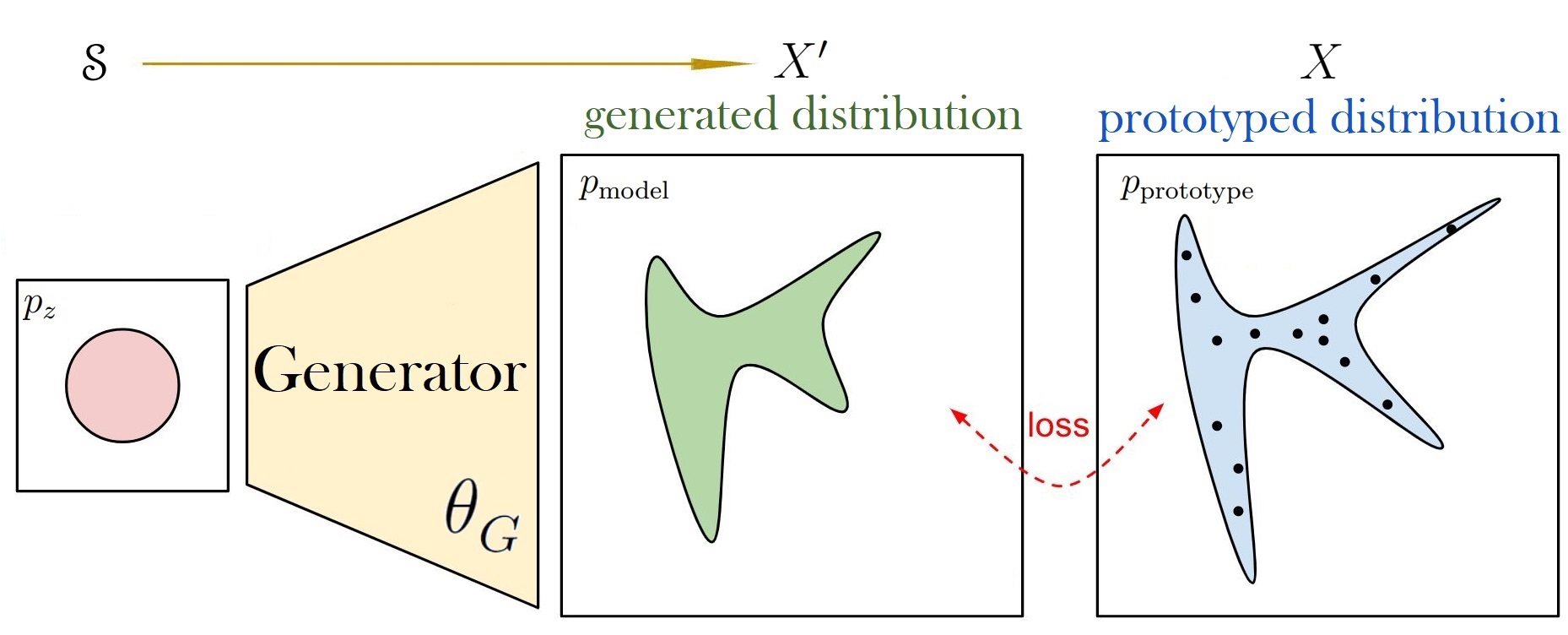}
	\caption{Global optimality formulation in probabilistic spaces \cite{open-ai-fig}}
	\label{fig:simple}
\end{figure} 

In a less formal illustration, in Fig.~\ref{fig:simple} \cite{open-ai-fig}, global optimality of the value function $V$ denotes convergence of the game. In more detail, the generator here (in yellow) learns in the best possible way, an implicit distribution \mbox{$p_\text{model}(x) \approx p_\text{prototype}(x)$} (respectively in green and blue) that inherently captures all prototyped dynamics of $X$ such as waveforms, channel effects and radio device imperfections. In the same figure, the space $\mathcal{S}$ can be viewed as the space of the $z$ samples (in red) transformed, at Nash equilibrium, to a space $X^\prime$ in a \emph{minimization} effort of a particular probabilistic loss function.

In deep generative modeling, the described game is called adversarial function approximation \cite{goodfellow-dl}, and the chained generator-discriminator DNN pair is called a \mbox{\emph{Generative Adversarial Network}} or GAN \cite{goodfellow-gan}. The adversary DNN players $D$ and $G$ are iteratively trained on stochastically optimizing $(\theta_D,\theta_G)$. This optimization is done using a supervised loss $\mathcal{L}_D$ (with labels on the two classes) for the discriminative model and an unsupervised embedded loss $\mathcal{L}_G$ for the generative model (hence the implicit density estimation notion), until a stop criterion, signalling convergence, is reached.
A training stop criterion makes the GAN chained model approximation subject to the failures of supervised learning: overfitting and underfitting \cite{goodfellow-dl,generative}. Other deep generative models make other approximations that are prone to other failures \cite{generative}. In principle, with rigorous optimization, hyper-parameter tuning and enough training data, GAN training limitations can be mitigated. The loss functions of the GAN model are explained in the following subsections.

\subsection{Loss function of the discriminative model}\label{disc-loss}
According to the original GAN paper \cite{goodfellow-gan}, the value function $V$ of both the GAN components can be written as a zero-sum game between two Kullback-Leibler divergences: where $D(x,\theta_D)$ is the discriminator output for prototyped data $x\sim p_\text{prototype}(x)$ and $D(G(z,\theta_G),\theta_D)$ is the discriminator output for generated data $G(z,\theta_G)$, i.e.:
\begin{equation}
    \begin{aligned}
     V(D,G)=&\mathbb{E}_{x \sim p_\text{prototype}(x)}[\log{}(D(x,\theta_D))]+\\
     &\mathbb{E}_{z\sim p_\text{z}(z)}[\log{}(1-D(G(z,\theta_G),\theta_D))].  \label{eq:value-function}
    \end{aligned}
\end{equation}

The quantity $V(D,G)$ can be viewed as the standard cross-entropy loss function of a binary classifier scaled by $-2$ where: data coming from the prototype is labeled “1” for all entries, and data coming from the generator is labeled “0” for all entries. The loss function of the discriminative model can then be written as
\begin{equation}
	\begin{aligned}
	\mathcal{L}_D(D,G)=-&\frac{1}{2}\mathbb{E}_{x \sim p_\text{prototype}(x)}[\log{}(D(x,\theta_D))]\\
	-&\frac{1}{2}\mathbb{E}_{z\sim p_\text{z}(z)}[\log{}(1-D(G(z,\theta_G),\theta_D))].
\end{aligned}\label{eq:l-d}
\end{equation}

\subsection{Loss function of the generative model}
Similarly in this adversarial training setting, the generator iteratively plays the zero-sum the game expressed by the value function (\ref{eq:value-function}). Therefore, the loss function of the generative model can be written as the opposite of the loss function of the discriminative model:
\begin{equation}
	\begin{aligned}
		\mathcal{L}_G(D,G)=&-\mathcal{L}_D(D,G);\\
		\mathcal{L}_G(D,G)=&\frac{1}{2}\mathbb{E}_{x \sim p_\text{prototype}(x)}[\log{}(D(x,\theta_D))]+\\&\frac{1}{2}\mathbb{E}_{z\sim p_\text{z}(z)}[\log{}(1-D(G(z,\theta_G),\theta_D))];\\
		\mathcal{L}_G(G)=&\frac{1}{2}\mathbb{E}_{z\sim p_\text{z}(z)}[\log{}(1-D(G(z,\theta_G),\theta_D))].
	\end{aligned} \label{eq:costG}
\end{equation}

The latter expression is useful for theoretical analysis. However, if we consider the case where the discriminator rejects the generator's samples with high confidence, we have:
\begin{equation}
	\begin{aligned}
		D(G(z,\theta_G),\theta_D) &\rightarrow 0;\\
		\mathcal{L}_G(G)=\frac{1}{2}\mathbb{E}_{z\sim p_\text{z}(z)}[\log{}(1-D(&G(z,\theta_G),\theta_D))] \rightarrow 0.
	\end{aligned} \label{vanish}
\end{equation}
Here, this situation indicates that the gradients on $\theta_G$ would vanish during training, i.e. the vanishing gradients problem. This would remain true unless a bias signal is added to the generator's parameters.

To solve this without biasing the prototype, one heuristically verified approach \cite{goodfellow-gan} is to continue to use cross-entropy minimization for the generator and instead of flipping the sign on the discriminator's loss function, the target used to construct the cross-entropy cost in \eqref{eq:value-function} is flipped, i.e.:
\begin{equation}
	\begin{aligned}
	\mathcal{L}_G(G) &= -\frac{1}{2}\mathbb{E}_{z\sim p_\text{z}(z)}[ \log{}(D(G(z,\theta_G),\theta_D))].\\ \label{eq:l-g}
\end{aligned}
\end{equation}
The generator would now try to maximize the log-probability of the discriminator being wrong. And when the latter rejects the former's samples with a high accuracy, gradients on $\theta_G$ also get better, i.e.:
\begin{equation}
	\begin{aligned}
		D(G(z,\theta_G),\theta_D) &\rightarrow 0;\\
		\mathcal{L}_G(G)=-\frac{1}{2}\mathbb{E}_{z\sim p_\text{z}(z)}[\log{}(D(&G(z,\theta_G),\theta_D))] \rightarrow \infty.
	\end{aligned}
\end{equation}
In this design, the game is no longer zero-sum, and it cannot be described with a single value function.

A good example to understand the vanishing gradients problem in the pseudo-radio-signal synthesis application is the differentiation of raw prototype data samples coming from an unknown wireless channel $H$. Broadly speaking, the discriminator does not make any assumptions on $p_\text{prototype}$ because it is a binary classifier initially trained in a supervised learning setting. Hence, $D$ is less prone to biases caused by channel agnosticism. Contrastingly, the generator's gradients can be blocked by the total blindness of the stochastic channel and the fact that its joint non-linearities are not modelled in $p_z(z)$ (in some pre-training version similar to $D$). In other words, when training the generator on minimizing the objective function (\ref{eq:value-function}), its gradients can turn out to be relatively flat (e.g. at high path-loss) and they would rapidly vanish as explained in (\ref{vanish}). This would make $G$ learn no inherent dynamic features while $D$ would gets too successful at its classification. Therefore, we use expression \eqref{eq:l-g} as the loss function of the generative model as a work-around to this issue.

\subsection{GAN advantages and challenges}
The last paragraph of the previous section depicts a good example on why domain knowledge is important in ML and DL. In fact, unless one is willing to make some assumptions about the problem from domain knowledge, estimating the optimal GAN parameters $(\theta_G^*,\theta_D^*)$ from the training set alone is evidently impossible no matter the training capacities. This impossibility is also known as the \mbox{\emph{no free-lunch theorem} \cite{SUR3}}, stating that: without making assumptions about the relationship between input and output, it is not possible to generalize the avaiable observations outside the training set. 

GANs potentially solve major limitations in learning end-to-end physical channel properties such as traffic pattern features, as we will show. We first note that the weights of the chained-GAN are usually updated using a stochastic gradient update algorithm that computes error gradients propagated from the discriminator to the generator. Using the prototyped input, the back-propagation of the gradients can be blocked (whitened) at the level of the discriminator or the generator when the channel $H$ is unknown or uncontrolled a priori. This situation prevents the overall learning of the end-to-end dynamics. Hence, the channel transfer function for example can be assumed and injected as a bias signal, but any such assumption would prejudice the learned weights, repeating the pitfalls caused by the discrepancy between the assumed and the actual channel like in simulators. In any case, in real wireless systems, an accurate estimate of $h(t)$ is usually hard to obtain in advance and it can not be expressed analytically. As a result, it is desirable to develop this channel agnostic traffic modeling tool of for pseudo-radio-signal synthesis, where different types of sandboxed waveforms and colored effects can be automatically learned without using analytical models.

In its essence, learning directly from complex systems with high degrees of freedom, such as radio hardware and wireless links, is generally troublesome \cite{GAN9}. Moreover, DL for wireless is a new field, and little is known about its optimal training strategies. In the rest of the paper, we aim to outline and experimentally solve some of the challenges (i.e. parameter tuning) in context of our GAN-based pseudo-radio-signal synthesis method.

\section{Pseudo-Radio-Signal Synthesis Method}\label{sec:prss-methodology}
\begin{figure}[t] 
	\centering
	\includegraphics[width=\columnwidth]{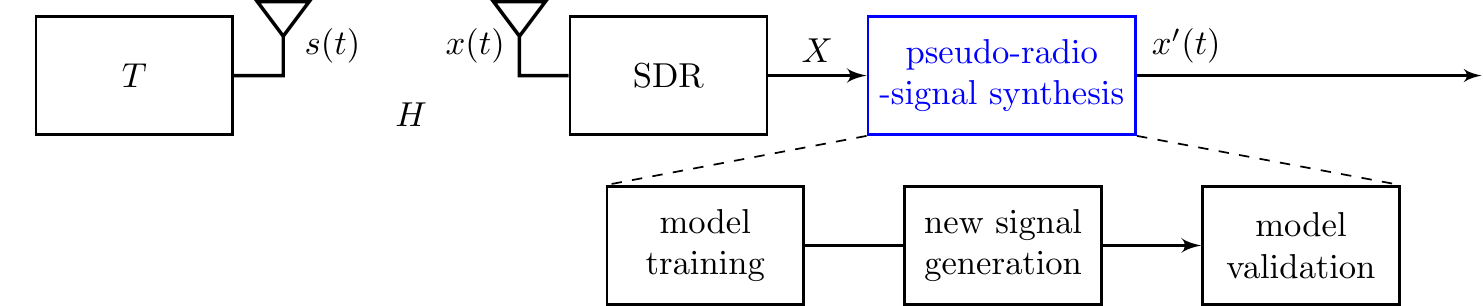}
	\caption{Block diagram staging the pseudo-radio-signal generation method}
	\label{fig:block3}
\end{figure}

We recall that the synthesis of a signal based on the prototype $x(t)$ yields a pseudo-radio-signal $x'(t)$ of duration $T_\text{gen}>T_\text{prototype}$. The GAN models prove to be lightweight (in order of few MBytes) and combine the prototyped waveforms, channel and hardware combined effects. This means that the analysis and engineering of wireless interference signals would only require information about the transmitter and the SDR hardware disposition in a known RF medium during {model training}. Systematically, this solves the storage problem by building, for different radio scenario compositions, a catalogue of agile models instead of raw signal recordings. The method will necessitate DSP and data science practices \cite{goodfellow-dl} invoking confidence metrics and performance testing on the model. For these reasons, a crucial model validation step is necessary before transmitting $x'(t)$ over-the-air. The general concept is illustrated in Fig.~\ref{fig:block3}.

On the other hand, with respect to the no-free-lunch theorem, few expert piloting is made in training and testing of the model. This includes:
\begin{itemize}
    \item the GAN model architecture,
	\item the prototype input data structure and pre-processing,
	\item The introduction of a signal-to-noise (SNR) variable in the latent prior $p_z$ statistics,
	\item regularizations and optimizations.
\end{itemize}

In the following, Section \ref{sec:ds} presents the framework and the data structure of the radio signals, following that, Section \ref{sec:norm} explains the normalization procedure prior to inputting the prototype into the model. Section \ref{model-training} explores the model training step for GAN and Section \ref{prss} exposes the necessary DSP for the synthesis of the new signal. Finally, Section \ref{sec:validation} details how we choose to validate the generated signal.

\subsection{Data structure}\label{sec:ds}
Real wireless signals are captured and transmitted in a controlled environment (i.e. RF anechoic chamber) using a Universal Software Radio Peripheral (USRP) \cite{x300-product} hardware in the SDR platform. The necessary DSP is implemented with Python \cite{uhd-python-api} and the DNN architectures are built using the Keras \cite{Keras} library. 

With an USRP gain equal to $G_\text{RX}$, the energy over time of any radio signal centred at $ f_\text{c} $ with a sampling bandwidth $ R_\text{s} $ is stored for the duration $T_\text{RX}$ in a vector array $s[n]$ having a number of samples $N_\text{s}$ holding the temporal in-phase and quadrature (I/Q) data components $a_n = I_n +{j} Q_n \in \mathbb{C}$ in a \verb|numpy.complex64| format and canonicalized as 
\begin{equation}
	\begin{aligned}
	s[n] &= a_n \in \Delta \mid \Delta = \{ \mathbb{C} : |a_n|=1 \}
	; n = 0 \dots N_\text{s}-1 ;\\
	N_\text{s} &=  \floor{T_\text{RX} \cdot R_\text{s}} \in \mathbb{N}_+.
	\end{aligned}
	\label{eq:s1}
\end{equation}

This I/Q complex-valued representation of the signal is commonly used in communications and is significant for many signal processing algorithms that rely on phase rotations, complex conjugates, absolute values, etc. For this reason, it would be desirable to have artificial neural networks operate on complex rather than real numbers. However, for several reasons, none of the available ML libraries, including Keras, supports this at the time of writing this paper. First, it is possible to represent all mathematical operations in the complex domain with purely real-valued neural network of twice the size, i.e., each complex number is simply represented by two real values. Second, a complication arises in complex-valued neural networks since traditional loss and activation functions are generally not holomorphic so that their gradient is not defined. A solution to this is Wirtinger calculus \cite{SUR4}. Although complex-valued neural networks might be easier to train and consume less memory, no research to this date has provided any significant advantage in terms of expressive power. For this reason, given a prototype $x(t)$ of duration $T_\text{prototype}$, its digital temporal representation must be reshaped in order to be used in the Keras environment outside of the USRP ecosystem. We define the tensor notation for a given prototype by
\begin{equation}
	T^x = c_{[d_1,d_2,d_3,d_4]} \in \mathbb{R}^{N_\text{f}^x \times N_\text{p}^x \times \text{DIM}_\text{IQ} \times N_\text{FFT}}.
\end{equation}
This 4D tensor is the input structure of the prototype for the proposed GAN implementation. $\text{DIM}_\text{IQ} = 2$ seperates the $I$ and $Q$ data components which downgrades the data structure of $c$ from \verb|numpy.complex64| to \verb|numpy.float32|. However, automatic differentiation environments for double valued neural networks are also not sufficiently mature, for the same previously stated reason that a double convolutional layer can obtain much of the benefit within this representation. consequently, it is believed that it is sufficient for the time being to use one $I$ or $Q$ component at a time. A number $N_\text{f}>0$ is used to construct different {frames} of the prototype in order to easily separate the training and the testing set as well as to  provide diversity in the pseudo-radio-signal synthesis process discussed in Section \ref{prss}. 
Each frame is composed of “packets” of a sequence of $N_\text{FFT}>0$ captured $c$ samples. The number of packets is equal to
\begin{equation}
N_\text{p}^x=\floor{\frac{N_\text{s}}{N^x_\text{f} N_\text{FFT}}} \in \mathbb{N}_+.
\end{equation}

\subsection{Normalization}\label{sec:norm}
Data normalization is an important step prior to any ML application \cite{goodfellow-dl}. Data values $\upsilon$ of each element $c$ of the prototype tensor $T^x$ are scaled to unit energy {for each frame envelop} $\psi$ using the linear transformation
\begin{equation}
		T^x_{[d_1=\psi,d_4=\upsilon]} \leftarrow  \frac{T^x_{[d_4=\upsilon]}}{\mathcal{P}_{\psi}};
\end{equation}
\begin{equation}
		\mathcal{P}_{\psi} = 
		\frac{1}{N_\text{p}^x}
		\sum\limits_{p=0}^{N_\text{p}-1}	
		\sum\limits_{\upsilon=0}^{N_\text{FFT}-1}
		d_{[d_1=\psi,d_2=p,d_4=\upsilon]};
\end{equation}
where $\mathcal{P}_{\psi}$ designates the average power of the frame $\psi$. This destroys any residual features which are simply real world artifacts possibly due to hardware (USRP or $T$) imperfections. Normalization also makes training of the GAN less sensitive to the scale of features and improves analysis and comparison of multiple models. In fact, normalization makes the data better conditioned for convergence because in case of massive variances, optimization can stagnate. This also allows to initialize the gradients $(\nabla_{\theta_D},\nabla_{\theta_G})$ to null values prior to training and keeps the weights $(\theta_D,\theta_G)$ values in $(-1,1)$.

\subsection{Model training}\label{model-training}
\begin{figure*}
	\centering
	\includegraphics[width=0.92\textwidth]{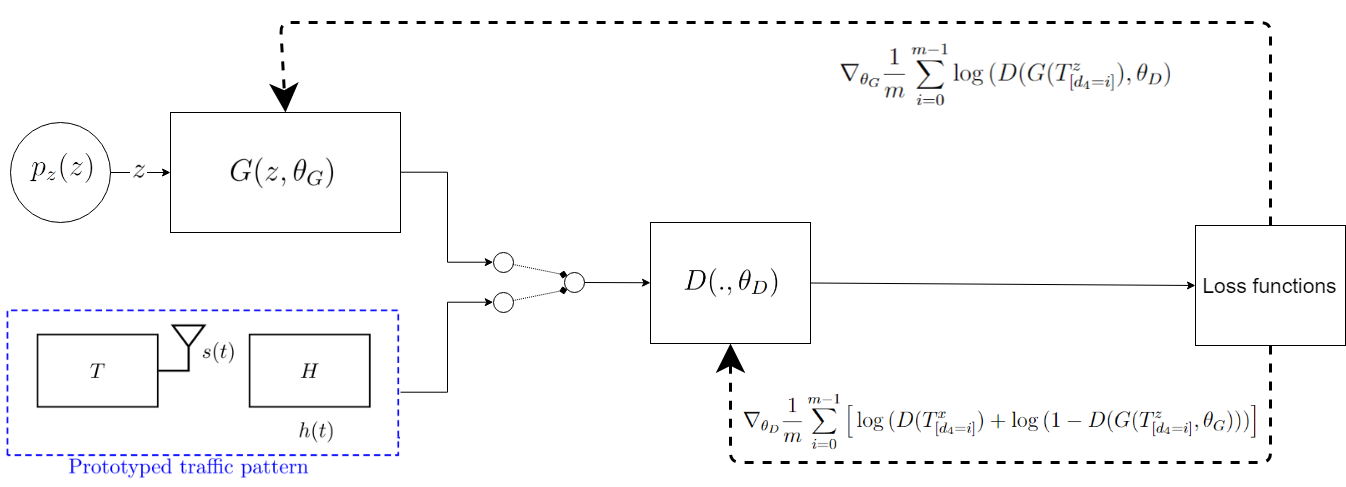}
	\caption{High level illustration of GAN model training for pseudo-radio-signal synthesis}
	\label{fig:draw-io}
\end{figure*}

To minimize the loss functions $\mathcal{L}_D(D)$ and $\mathcal{L}_G(G)$ expressed respectively in the equations (\ref{eq:l-d}) and (\ref{eq:l-g}), the feed-forward back-propagation algorithm \cite{GSN} with mini-batch \cite{goodfellow-dl} is used. Feed-forward is the algorithm that calculates the output vector from the input vector of each layer of a DNN. Back-propagation is the algorithm used to stochastically update the weight parameters $(\theta_D,\theta_G)$ using the respective gradients $(\nabla_{\theta_D},\nabla_{\theta_G})$ according to an optimizer. Fig.~\ref{fig:draw-io} illustrates this within the pseudo-radio-signal synthesis framework.

The chained-GAN has two modes: the training mode (for updating the weights $(\nabla_{\theta_D},\nabla_{\theta_G})$) and the generation mode (for pseudo-radio-signal synthesis). In the generation mode only feed-forward for $G$ is used, while in the training mode, both algorithms are used.

In the training mode, $D$ and $G$ are trained iteratively on the the training set which is {one} ($d_3$) $\text{DIM}_\text{IQ}$ dimension of {one} ($d_1$) frame with $N_\text{examples}<N_\text{f}^x$ packets from $T^x$ and $N_\text{examples}$ packets from $G(z;\theta_G)$. Both example entries are randomly fed to the discriminator (with the corresponding binary classification labels) a number of $N_\text{epoch}$ iteration times. The tensor data values are fed by chunks of size  $S_\text{mini-batch}<N_\text{FFT}$ in order to parallelise and optimize the learning process. Embedded updates on $G(z,\theta_G)$ happen with the feed-back sent from the discriminator as highlighted in Fig.~\ref{fig:draw-io}.

Initialization of the two DNN weights has no theoretical foundations for GAN at the time of writing this work. For both the discriminator and the generator, we chose to utilize the practical technique of Xavier initialization \cite{xavier}. The discriminator's weights are initialized in an earlier number of epochs $N_\text{epoch-pretrain}$ for pre-training. In principle, with a sufficient stop criterion $N_\text{epoch}$, the generator is observed to be able to approximate dynamics of a prototyped input in terms of packets. The output pseudo-radio-packets are tested in the validation step explained in Section \ref{sec:validation}.

The $z$ values are sampled from $p_z(z)$ which can be a uniform distribution or a Gaussian distribution. The choice of the $p_z(z)$ density is critical and it is an active research topic \cite{sampling} in the DL community. In our work, we propose to use a Gaussian distribution for the latent noise with mean $\mu$ and variance $\sigma^2$, i.e. $z \sim \mathcal{N}(\mu,\sigma^2)$. Consider the training step given the prototyped frame ${\psi}$. The proposed $z$ introduces favorable gradient propagation at the generator weights to simulate wireless effects: with a noise power level $\sigma^2$ corresponding to a signal-to-noise (SNR) ratio $\gamma_\text{dB}$ of the known average power $\mathcal{P}_{\psi}$ w.r.t. the noise $z$.
   
With sufficient training capacities, using Algorithm \ref{alg:sigma}, and for an adequate SNR value $\gamma_\text{dB}$ at each epoch, we observe that the latent noise $z\sim\mathcal{N}(\mu,\Lambda(\mathcal{P},\gamma_\text{dB}))$ can virtualize additive channel effects like noise and co-channel and adjacent channel interferences existent in the prototyped waveform.
   
    \begin{algorithm}[b]
    	\text{~~~~\textbf{input:  }Signal power level $\mathcal{P}$ and SNR ratio $\gamma_\text{dB}$}\\
    	\text{~~~~\textbf{output: }Noise power level $\sigma^2$ }\\
    	\text{~~~~function $\Lambda$ $(\mathcal{P},\gamma_\text{dB})$:}\\
    	$~~~~~~\mathcal{P}_\text{dB}=10\times\log_{10}  (\mathcal{P})\;$\\
    	$~~~~~~\text{noise}_\text{dB}=\mathcal{P}_\text{dB}-\gamma_\text{dB}$\;\\
    	$~~~~~~\sigma^2=10^{\frac{\sigma_\text{dB}}{10}}$\;
    	\caption{Pseudo-code to compute latent Gaussian noise variance from virtual SNR quantity.}\label{alg:sigma}
    \end{algorithm}
   
For robust feature extraction, we highlight that the mini-batch size during training $S_\text{batch}$ should be few-orders bigger than the smallest kernel or dense layer dimensions utilized in the chained-GAN and at the same time few-orders smaller than the packet size, i.e. $\text{min}(S_\text{k},\zeta_D,\zeta_G) \ll S_\text{batch} \ll N_\text{FFT}$. In practical wireless communication environments, this would be useful for radio channel feature learning over a variety of stochastic effects and also for pseudo-radio-signals synthesis.

It is not obvious, however, at which SNRs the mini-batches should be trained. It is clearly desirable that the GAN should operate at any SNR or SNR-range. However, previous research argues that this is generally not the case. In \cite{R2}, training at low SNR did not allow for the discovery of the wanted structure important in higher SNR scenarios. Also, training across a wide range of SNR severely affects training time. On another hand, authors of \cite{snr-epoch} have observed that starting off the training at high SNR and then gradually lowering it with each epoch led to significant performance improvements for their DL application in astrophysics. In our proposed solution, for each prototype $X$, and enabled by normalization, the $\mu$ value for the latent Gaussian distribution is set to $0$ and $\sigma^2=\Lambda(\mathcal{P},\gamma_\text{dB}^x)$ is chosen at each epoch with $\gamma_\text{dB}^x \in \Gamma^x=[\gamma_\text{dB,min}^x \dots \gamma_\text{dB,max}^x]$.

For a fixed scenario and a fixed USRP receive gain $G_\text{RX}$ (i.e. in a controlled environment), results have heuristically shown that, with convenient and thorough regularizations (to avoid overfitting and underfitting) on the GAN architecture, it is {best-practice} to randomly pick SNR values $\gamma_\text{dB}^x$ from $\Gamma^x$ uniformally at each epoch. This way, $G(z,\theta_G)$ builds a probabilistic generative model $p_\text{model}(x)$ at the end of the training mode. The reasoning behind this is that the GAN learns more radio dynamics and channel effects for different noise level variances at each epoch. This can be viewed as a special kind of regularization itself: since in training mode, picking different SNRs can be extended to other operational situations. For example, with the same GAN architecture, deriving SNR-range profiles $\Gamma^x$, based on other digitally simulatable radio parameters, enables the building generative models for more complex traffic patterns, e.g., including transmitter mobility or a harsher RF propagation environment.

\subsection{New signal synthesis}\label{prss}
At stop criterion, in the generation mode, associated neural outputs of the output layer of the generator are able to construct a current tensor output $G_{N_\text{epoch}}(T^z)=T^{x^\prime}$ holding a frame of pseudo-radio-packets of size $N_\text{FFT}$ generated using the eventually learned and detected patterns of the input. $G$ is able to generate any arbitrary number $N_\text{gen}$ of these packets.

Note that the frames of the prototype have a power statistic $\mathcal{P}$. Therefore, randomizing the frame identifier $d_1$ ($N_\text{examples}$ times) in a range less than $N_\text{f}$ diversifies some generated packets. This allows us to build the test-set after rescaling these output frames (from $G$) for denormalization.

Doing this process in both $\text{DIM}_\text{IQ}$ dimension gives $N_\text{gen}^I$ and $N_\text{gen}^Q$ pseudo-sequences. These are assembled in the I/Q format in a 2D matrix $\mathbf{A}^{x^\prime}_{[p,\tau]} \in \Delta^{N_\text{gen} \times N_\text{FFT}}$. As we show next, visualizing this pseudo-radio-packets matrix $\mathbf{A}^{x^\prime}$ reveals serious phase shifts because the $N_\text{FFT}$ packet size is not necessarily chosen using any a priori information about the traffic. This means that the packet boundaries can reflect desynchronizations between the USRP and the transmitter learned as features during training. Such phenomena can be viewed as unfavorable co-channel and adjacent channel interference. This can be overcome, using the Griffin-Lim \cite{g-l} method commonly used in audio reconstruction, or by adding bias on the GAN while training using the same Griffin-Lim method \cite{GAN11}. It is also possible to train a separate GAN to learn the phase patterns, and any combination between GANs and the Griffin-Lim method is attractive for future research.

Herein, reconstruction is made using a discrete circular convolution ($\oast$) on the time domain matrix $\mathbf{A}^{x^\prime}$ with a discrete raised-cosine (RC) filter $\text{h}_\text{RC}[n]$, i.e.:

\begin{equation}
	x'[\tau] = \mathbf{A}^{x^\prime}_{[p,\tau]} \oast_{N_\text{FFT}}{h}_\text{RC}[\tau].
\end{equation}
\begin{equation}
	\text{h}_\text{RC}[\tau,L,\beta]=
	\begin{cases*}
		1,\phantom{aaaaaaaaaaaaaaaaaaa} |\tau| \leq \frac{1-\beta}{2L} &\\
		\frac{1}{2}\Big[1+\cos\left(\frac{\pi L}{\beta}(|\tau|-\frac{1-\beta}{2L})\right)\Big], \\ \phantom{aaaaaaaaaaaaaaaaaaaaa} \frac{1-\beta}{2L}<|\tau|\leq\frac{1+\beta}{2L}& \\
		0,\phantom{aaaaaaaaaaaaaaaaaaa} \text{otherwise.} &
	\end{cases*} 
\end{equation} \label{eq:rc}

Filtering is done between consequent packets using an overlap-save technique with padding \cite{overlap-save}. The chosen filter length is $L=129$ with the roll-off $\beta$ equal to $0.25$. Other values give other boundary reconstruction qualities.

\subsection{Model validation}\label{sec:validation}
Given that the prototype is totally arbitrary, it is necessary to choose a sandbox set of fixed inputs and assess the GAN performance on them as proof of concept. This is called model validation, and is done to evaluate the spectral and temporal radio properties of the synthesized pseudo-radio-signal. For this we use the same frames utilized in the training mode to construct the validation-set in the temporal domain.

Discrete Fourier transform using \verb|numpy.fft.fft| \cite{numpy} is applied on the prototype's 2D temporal representation $\mathbf{A}^x$, to yield a matrix $\mathbf{B}^x = b_{[\nu,p]} \in \Delta^{ N_\text{FFT} \times N_\text{p}}$ holding frequency samples $b_m$ with the finite-precision duration $N_\text{FFT}$. For sake of generality, and to insure that the GAN learns relevant frequency dynamics while keeping diversity, the spectral notation of the prototype $\mathbf{B}^x$ is compared to the output of the generator for an equal number of packets. The goal is to test the GAN's ability to learns qualitative spectral features.

On another hand, The GAN temporal output will be quantitatively compared to the prototype using their probabilistic density functions (PDF) $F_{p_\text{model}}$ and $F_{p_\text{prototype}}$ respectively. This is a classic approach for comparing the time varying behaviors of signals. Alternatively, the dependency on the time-series samples of the prototype can be addressed using recurrent neural network(s) sequence modeling. This topic is interesting but is out of our research scope here.

\section{Proposed GAN Model}\label{sec:proposed}
In this section we introduce our proposed GAN model and its implementation using the previously presented tensor data structure. The model includes optimization techniques for computational efficiency in terms of training time and memory occupation, as well as regularization \cite{goodfellow-dl} techniques to avoid situations where the generator's output is equivalent to a random noise (underfitting) and to avoid reusing the same pseudo-radio-packet in synthesis (overfitting). 

\subsection{Loss functions and optimizers implementations}\label{sec:implementation}
Inside Keras, different loss functions can be used for adversarial function approximation problems, which lead to different variants of GANs. Herein, we choose to continue with the simplest approach as advised in the litterature \cite{goodfellow-dl}. We implement the loss functions elaborated in Section~\ref{sec:gan0}.  

At the time of writing this paper, little is known about hypothetical GAN parametrization and training practices, however a literature of heuristics exists \cite{hard0}. The adaptive moment estimator (Adam) \cite{adam} is chosen as the optimizer \cite{optimizers} for both the discriminator and the generator as it gives the best simulated validation results. The same choice manifests in other (DL and) GAN applications and variants such as deep convolutional generative adversarial networks \cite{DCGAN}. It is worth to stress that this depends on the application and other degrees of freedom. For example, the original GAN paper \cite{goodfellow-gan} used a momentum optimizer \cite{optimizers}. The same philosophy applies to the rest of the hyper-parameters.

We detail in the sequel the tuned architectures of the proposed DNN models and their respective hyper-parameters verified according to the model validation methodology described in Section \ref{sec:validation}.

\subsection{The generative DNN model}
Table \ref{tab:hiddenG} and table \ref{tab:hypG} both summarize the hyper-parameters of the fully connected DNN of the generative model $G(z;\theta_G)$.

\begin{table}[ht] 
	\centering
	\caption{Layers of the generative DNN model}
	\begin{tabular}{p{0.26cm}p{1cm}p{2.2cm}p{3.8cm}}
		
		\toprule

		\#&\text{Layer type} &{Parameters} &{Description}\\
		\midrule
		L0&		Input layer			&$ S_\text{L0}=N_\text{FFT}$ 					&
		A tensor holding a packet of size $N_\text{FFT}$ $z$ values sampled from $p_z(z)$\\
		\hline
		L1&	Non-linear dense layer &$a_1(\cdot)=\tanh(\cdot)$& Fully connected layer with a non-linear activation function $a_{1}(\cdot)=\text{tanh}(\cdot) \rightarrow (-1;1)$ applied on L0's weighted output.\\
		\hline
		L2&Non-linear dense layer w/ weight decay & $a_2(.)=\tanh(.)$ $\lambda_g=0.001$ & Fully connected dense layer with L1's weighted output with weight decay.\\
		\hline
		L3&		Output layer			&$ S_\text{L0}=N_\text{FFT}$ 					&
		A tensor holding a packet of size $N_\text{FFT}$ with values equal to $G(z,\theta_G)$\\
		\bottomrule
	\end{tabular}
	\label{tab:hiddenG}
\end{table} 

\begin{table}[ht] 
	\centering
	\caption{Generative model hyper-parameters}
	\begin{tabular}{p{3.9cm}p{4.1cm}}
		\toprule
		\text{Hyper-parameter} &{Value}\\
		\midrule
		Stochastic optimizer of the generator	&  Adam      ($\eta_G,\beta_1=0.9,\beta_2=0.999$)  \\
		\hline
		Progressive learning rate	& $\eta_G = 0.011$  \\
		\hline
		Dense layers dimension &  $\zeta_G=128$ \\
		\bottomrule
	\end{tabular}
	\label{tab:hypG}
\end{table} 

According to \cite{GAN8} and \cite{goodfellow-dl}, normalization is a key step enabling the generative model to converge using the non-linear activation function $\tanh$, usually used in the generator's first and last layers. It is also worth noting that using dropout layers in the generative model would cause underfitting in the resulted pseudo-radio-packets.

For each prototype, the tensor notation allows the chained-GAN model to train using  $2\times N_\text{examples}$ frames of size $N_\text{p}$ of packets of size $N_\text{FFT}$. Half of these packets come from the generator's output layer and the other half comes from the prototype training frame as explained in Section \ref{model-training}.

\subsection{The discriminative DNN model}

The discriminative model $D(\cdot;\theta_D)$ is a fully connected DNN binary classifier with the tuned hyper-parameters presented in both table \ref{tab:hiddenD} and table \ref{tab:hypD}.

\begin{table}[ht] 
	\centering
	\caption{Layers of the discriminative DNN model}
	\begin{tabular}{p{0.26cm}p{1cm}p{2.2cm}p{3.8cm}}
		\toprule

		\#&\text{Layer type} &{Parameters} &{Description}\\
		\midrule
		L0& Input layer&$ S_\text{L0}=N_\text{FFT}$	& 
		Normalized input (packets) of size $S_\text{L0}$ from either the prototype $T_\text{X}$ or $G(z;\theta_G)$.  \\
		\hline
		L1&1D convolution layer & $N_\text{k}=32\phantom {aaaaaaa} \linebreak S_\text{k}=128$ & 
			$N_\text{k}$ filters (\text{kernels}) of size $S_\text{k}$ producing $S_\text{k}$ features from L0. Output of L1 is a 5D tensor with a dimension of size $S_\text{k}$ holding weights for each filter.\\
		\hline
		L2&	Non-linear dense layer & $a_2(\cdot)=\text{ReLu}(\cdot)$
		Rectified Linear Unit& Fully connected layer with a non-linear activation function $a_2(\cdot)=\text{ReLU}(\cdot)={max}(0,\cdot) \rightarrow [0;1]$ applied on L1's weighted output. \small{(max($\cdot$)=1 by reason of normalization.)}\\
		\hline
		L3&Dropout layer & $\delta_D=0.5$ & Is a regularization technique where weight connections between L2 and L3 are randomly dropped out with a probability of $\delta_D$ to ensure the same feature is learned with different connection architectures during training \cite{goodfellow-dl}.\\
		\hline
		L4& Flatten layer &  & Consecutively joins L3 weights (separated by L1) in a 4D tensor.\\
		\hline
		L5& Dense layer w/ weight decay & $\lambda_D=0.0001 $ & Fully connected layer (with a linear activation function $a_5(\cdot)=\cdot$) connecting L4's weighted output with rigid penalization on its output weights done by adding a factor equal to $\lambda_D||\mathbf{W}||^2_2$ to the loss function. This is called the weight decay regularization technique and it is useful to avoid overfitting for unbiased classifications over large neural networks \cite{goodfellow-dl}.\\
		\hline
		L6&Dropout layer & $\delta_D=0.5$& Dropout layer on L5.\\
		\hline
		L7&Dense layer w/ weight decay & $\lambda_D=0.0001 $ & Fully connected layer with weight decay on L6.\\
		\hline
		L8&Dropout layer & $\delta_D=0.5$ & Dropout layer on L7.\\
		\hline
		L9&Dense layer&  & Dense layer with with L8.\\
		\hline
		L10&Dropout layer & $\delta_D=0.5$ & Dropout layer on L9.\\
		\hline
		L11&	Non-linear dense layer &$a_{11}(.)= \phantom {aaaaa} \linebreak{softmax}(.)$& Fully connected layer with a non-linear activation function $a_{11}(\cdot)=\text{softmax}(\cdot)= \frac{e^{(z_i)}}{\sum_j \exp{(z_j)}} \rightarrow (0;1)$ applied on L10's weighted output.\\
		\hline
		L12 &		Output layer & $S_\text{L12}=2$ 					&
		Vector containing probabilities of the input packet belonging to each class, that is, probability of i) belonging to the prototype $T^x$, and probability of ii) being generated by $G$.\\
		\bottomrule
	\end{tabular}
	\label{tab:hiddenD}
\end{table} 

\begin{table}[ht] 
	\centering
	\caption{Disciriminative model hyper-parameters}
    \begin{tabular}{p{3.8cm}p{4.1cm}}
		\toprule
		\text{Hyper-parameter} &{Value} \\
		\midrule
		Stochastic optimizer of the discriminator		&  Adam      ($\eta_D,\beta_1=0.9,\beta_2=0.999$)  \\
		\hline
		Progressive learning rate	& $\eta_D = 0.0001$  \\
		\hline
		Binary classification labels 		& $1-\alpha$ for the prototype examples. $\alpha=0.2$\\
		&0 for the generator's output.\\
		\hline
		Dense layers dimension&  $\zeta_D=32$ \\
		\hline
		$N_\text{epoch-pretrain}$&  1 \\
		\hline
		$S_\text{mini-batch-pretrain}$& 32 \\
		\bottomrule
	\end{tabular}
	\label{tab:hypD}
\end{table} 

Some individual packets from the prototype can include negative clipping effects caused either by the USRP's analog-to-digital converters or sharing the channel. These can escape the normalization constraints and can cause un-helpful gradients at some epochs. This manifests as sequences with very low scalar data values compared to previous epochs. Moreover, this situation can cause saturation during training especially if the the generator is outputting large pseudo-radio-packets at the time. In this game, the discriminator (which is a classifier) tends to linearly extrapolate and produce extremely confident predictions that cause the saturation. If $p_\text{prototype}(x) \rightarrow 0$ and $p_G(x)$ is large, erroneous samples from the current $p_G(x)$ then have no incentive to approximate the data.

To encourage the discriminator to estimate soft probabilities rather than to extrapolate to extremely confident classifications, the regularization technique of \mbox{\emph{one-sided label smoothing}} \cite{hard0} is used. The parameter $\alpha$ is the smoothing parameter. It is important to smooth only the prototyped data \cite{generative}, which explains the term “one-sided”. In fact, non-zero values for generated data coinciding with a large $p_G(x)$ and a small $p_\text{prototype}(x)$ would reinforce incorrect behavior in the generator. This would train $G$ either to produce samples that resemble the data or to produce samples that resemble the sequences it already makes.

\subsection{The chained-GAN model}
The constants in table \ref{tab:chained} construct the set of hyper-parameters for the chained-GAN model (see Fig.~\ref{fig:draw-io}) which subsequently determine the number of overall trainable parameters $(\theta_G,\theta_D)$ and final model size. These values are chosen for big enough frames to show validation results in a timely manner using TensorFlow's \cite{tensorflow} central processing unit (CPU) backend for various prototypes.

\begin{table}[ht] 
	\centering
	\caption{Chained-GAN model hyper-parameters}
	\begin{tabular}{lp{6cm}}
		\toprule
		\text{Hyper-parameter} &{Value}\\
		\midrule
		$N_\text{examples}$	&  128  \\
		\hline
		$N_\text{FFT}$	&  2048  \\
		\hline
		$N_\text{p}$	&  $\sim 400$: kept as constant for different prototypes and chosen w.r.t. dividing the smallest validation prototype to a number of frames $N_\text{f}=2$  \\
		\hline
		$N_\text{gen}$& $N_\text{p} \times 20$ \\
		\hline
		$N_\text{epoch}$&  1000 \\
		\hline
		$S_\text{batch}$& 300 \\
		\bottomrule
	\end{tabular}
	\label{tab:chained}
\end{table} 

So far, the stop criterion is the number of epochs $N_\text{epoch}$, which theoretically gives better results the more the chained-GAN trains. However, given the limited amount of prototyped data (due to wideband signal recording), longer training can conducive to overfitting. For this reason, the training losses and the discriminator's accuracy \cite{Keras} are manually monitored while tuning the model. The monitoring of these single valued metrics resulted in valuable preliminary observations in this application. Advanced early stopping \cite{goodfellow-dl} based on optimal model weights is encouraged to be investigated in the future.

\section{Numerical Results}\label{sec:numerical-res}
In this section, we present training and validation numerical results for our method illustrated in Fig.~\ref{fig:block3}. Additionally, we discuss insight and remarks.

The following example is based on a real signal processed via a USRP X300 \cite{x300-product} SDR for prototyping and transmission of pseudo-radio-signals. The selected prototype $x[n]$ is a wireless signal recording of duration $T_\text{RX}=\SI{3}{s}$ and an $ R_\text{s} =\SI{200}{MHz} $ sampling band centred at $ f_\text{c} = \SI{5.25}{GHz}$ and captured with a receive gain $G_\text{RX}=\SI{31.5}{dB}$ in a controlled environment. This is the unlicensed WiFi channel 50 where we generate the traffic at maximum benchmark transfer rate using an IEEE 802.11ax capable transmitter $T$ at a distance equal to \SI{4}{m} from the USRP.

\subsection{Training of the proposed model}\label{num:training}

During training, the mean success rate across all predictions for binary cross-entropy problems can be referred to as accuracy inside of Keras. The optimal accuracy for the discriminator for each training process is equal to 0.5 as explained in Section \ref{sec:afa}. If the accuracy is more than 0.5 then the GAN output is exposed to overfitting; if the accuracy is less than 0.5, then the GAN output is exposed to underfitting.

\begin{figure}[b]
	\centering
	\includegraphics[width=0.89\columnwidth]{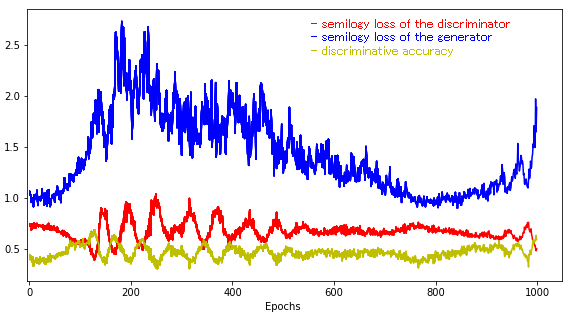}
	\caption{Training metrics evolution on the prototype}
	\label{fig:training}
\end{figure}

Fig.~\ref{fig:training} jointly shows the discriminative loss (in red), the generative loss (in blue) and the discriminator's accuracy (in yellow), during an $\SI{8}{mn}:\SI{38}{s}$ training of $N_\text{epochs}=1000$ on $T^x$ with a decibel SNR-range $\Gamma^x=[-30 \dots -24]$.

After trial with different types of prototype signals, the most important notes for the GAN training mode, that happen to be common in most GAN applications are the following:

\begin{itemize}
	\item The chained model's architecture and hyper-parameters are evidently key for training convergence. In this example, and for the mentioned duration of training, the average discriminative accuracy is $0.5898$, which implies chances of overfitting. The accuracy metric is important, because training can continue even if the accuracy is saturated, which implies that the model is defective. In such a case, divergence of the chained model is inevitable, and saturation of the accuracy is an indicator that $N_\text{epochs}$ must be reduced. 
	\item A model with losses converging very fast to zero is defective. A model with discriminative losses having huge variances and spikes is defective. A model with generative losses steadily increasing is defective. These unbalances obviously indicate underfitting or overfitting.
	\item The loss function of the discriminator and the loss function of the generator are opposite (adversarial) and the amount of times each model gets better relatively to the other is interesting to monitor to compare different chained-GAN architectures. High absolute peaks in the relative losses imply changes in the SNR value and subsequently allowing the GAN to learn new radio patterns.
\end{itemize}

\subsection{Validation of the proposed model}\label{num:validation}
\begin{figure}[ht]
	\centering
	\includegraphics[width=0.85\columnwidth]{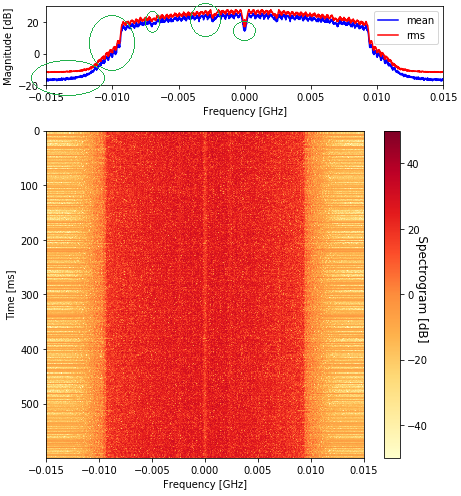}
	\caption{Visualization of spectral features of a sampled signal from prototype}
	\label{fig:sampled}
\end{figure}

\begin{figure}[ht]
	\centering
	\includegraphics[width=0.85\columnwidth]{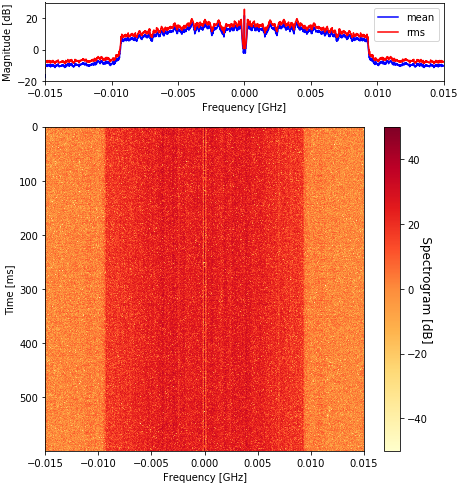}
	\caption{Visualization of spectral features of a generated signal by the generator}
	\label{fig:gen}
\end{figure}

Let us first evaluate the spectral dynamics, i.e. $\mathbf{B}$. Fig.~\ref{fig:sampled} shows visualization of $N_\text{gen}$ packets samples from $T^x$. Fig.~\ref{fig:gen} shows visualization of $N_\text{gen}$ pseudo-radio-packets sampled from $G_{N_\text{epoch}}(T^z)=T^{x^\prime}$ of the proposed GAN model in its generation mode.

The generator's output (Fig.~\ref{fig:gen}) is colored and is different from a Gaussian noise. This indicates that $G$ has the learned qualitative degrees of the spectral dynamics (circled in green in Fig.~\ref{fig:sampled}) of the prototyped signal. The output quality is linked to the model architecture and hyper-parameters. For different regularization techniques, table \ref{tab:comparison} shows performance comparison using the same prototype and $N_\text{epochs}$. In particular, it is worth to note that the most influential regularization technique is one-sided label smoothing, which allowed to learn the magnitude peaks efficiently. Pay-off for different values of $\alpha$ is an open quest for better performance.
\begin{table}[ht] 
	\centering
	\caption{Performance comparision using different regularization techniques}
	\begin{tabular}{lcc}
		\toprule
		\text{Regularization} &{Average $D$ accuracy} &{Runtime} \\
		\midrule
		No regularization & 0.1822 & $\SI{22}{mn}:\SI{30}{s}$ \\
		\hline
		Dropout & 0.3001 & $\SI{10}{mn}:\SI{04}{s}$ \\
		\hline
		Weight decay & 0.6652 & $\SI{32}{mn}:\SI{40}{s}$ \\
		\hline
		One-sided label smoothing & 0.7699& $\SI{26}{mn}:\SI{10}{s}$ \\
		\bottomrule
	\end{tabular}
	\label{tab:comparison}
\end{table} 

Moreover, $p_\text{prototype}$ is dubbed multi-modal, meaning that many modes (of selectable packets) are possible. For non-convex games, this can cause the gradients to not converge using gradient descent optimizers (see Section \ref{sec:implementation}). This is because, intuitively, the longer the training, the harder it is for the DNN models to converge into describing all modes (or radio dynamics). Mathematically, this makes it harder for the chained model to converge using the cross-entropy loss functions to describe the adversarial game because of the non-symmetry of the Kullback-Leibler divergence (which drops to 0 for areas where $p_\text{prototype}(x) \rightarrow 0$). As discussed in Section \ref{sec:gan0}, this makes some gradients vanish during propagation as we discussed. Such flat gradients are common issue for GAN known as \emph{mode-collapse} \cite{9312049} which states that: there is inherently no motivation for the generator to produce a diverse set of samples because the discriminator only penalizes for producing bad samples. Therefore, it easier for the generator to learn a few modes than all modes for a multi-modal distribution  as we see in Fig.~\ref{fig:gen}.

\begin{figure}[ht]
	\centering
	\includegraphics[width=0.9\columnwidth]{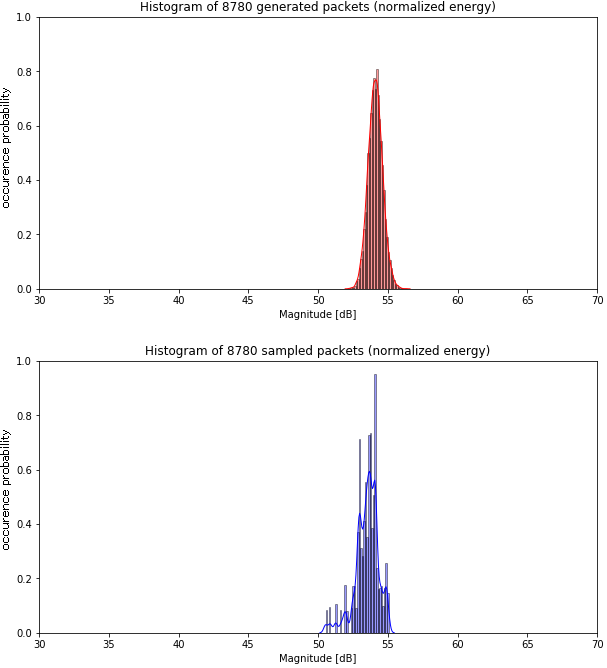}
	\caption{PDFs of signals from both the prototype and the generator's output}
	\label{fig:pdf}
\end{figure}

\begin{figure}[ht]
		\centering
		\includegraphics[width=0.75\columnwidth]{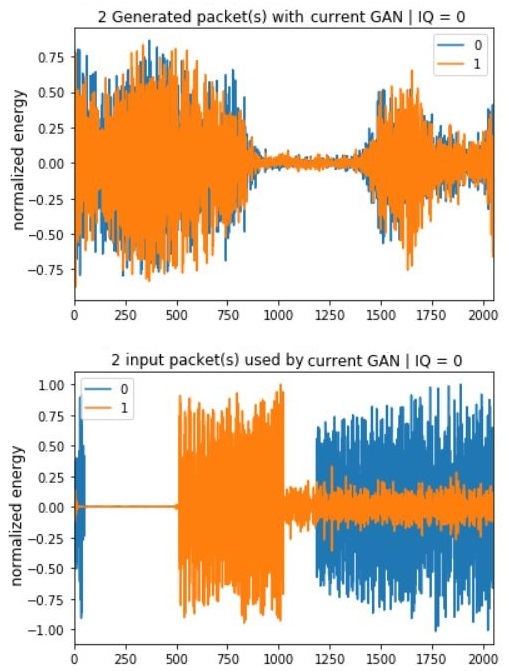}
		\caption{Two packets from both the prototype and the generator's output}
		\label{fig:4p}
    \end{figure}

Moving on to the temporal dynamics, i.e. $\mathbf{A}$, Fig.~\ref{fig:pdf} shows the PDFs of the previously presented signals and the quantitative time-domain performance. In particular, Fig.~\ref{fig:4p} shows two generated pseudo-radio-packets and two randomly sampled packets of length $N_\text{FFT}=2048$.

Classically, working with RF signals numerically transformed to stochastic time-series has always brought into attention the time-slotted property of the signal; meaning that the signal's variance (and basically it's power) depends on the time of measurement. Some simulation models assume all disturbances to be stationary while others are built around the specific notion that interference is not stationary. For the GAN to model complex traffic models (especially with a transmitter in movement, more dense environment, fading, etc), the generator would find it difficult to map the traffic with a fixed $N_\text{FFT}$. If $N_\text{FFT}$ is a trainable degree of freedom, the generator could potentially learn the radio channel response and synchronize with the actual prototype data symbols. In this case, cyclic stationarity becomes important in training. This can be fixed with a cyclic correlation logic \cite{principles2}. Namely, if the training size $N_\text{p} \times N_\text{FFT}$ is close to some cyclic window of the traffic, GAN would converge faster. 
This could be addressed using recurrent neural network(s) sequence modeling as a dependency on the time-series samples of the prototype or using another GAN to help find an adequate $N_\text{FFT}$. This bias can then be added to the first GAN each time it is necessary. Alternatively, for a fixed $N_\text{FFT}$, a lower DNN dimensions $\zeta_D$ and $\zeta_G$ can be used alongside a higher training time.

\section{Conclusion}\label{sec:conclusion}
In this work, we proposed a DL-based channel agnostic lightweight traffic modeling tool. Different types of traffic patterns and channel effects can be automatically learned without knowing details about the prototyped traffic or the channel transfer function.

By using an adversarial approach, we showed that it is possible to learn function approximations for arbitrary over-the-air communication traffics with colored characteristics such as traffic waveform, noise and channel interferences. For our pseudo-radio-signal synthesis method, we proposed a GAN model featuring an implicit regularization technique to detect spectral and temporal dynamics of the prototype around a range of virtual SNR(s).

GANs and ML in general, come with the promise of potential gains over existent algorithms for solving non-linear problems for the physical layer. While enhancements are yet possible for the proposed model in terms of improving stability and performance, we have, nonetheless illustrated that such a DL approach, can, even with a relatively simple architecture, obtain reasonable traffic approximations without the introduction of assumptions about the effects occurring or the simplification to a parametric model.

\section*{Acknowledgments}
This work is partially supported by the Luxembourg National Research Fund (FNR) under the frame of the research project (BRIDGES19/IS/13778945/DISBuS).
 
\bibliographystyle{IEEEtran}
\bibliography{IEEEabrv.bib,GAN_bib.bib}

\begin{IEEEbiography}[{\includegraphics[width=1in,height=1.25in,clip,keepaspectratio]{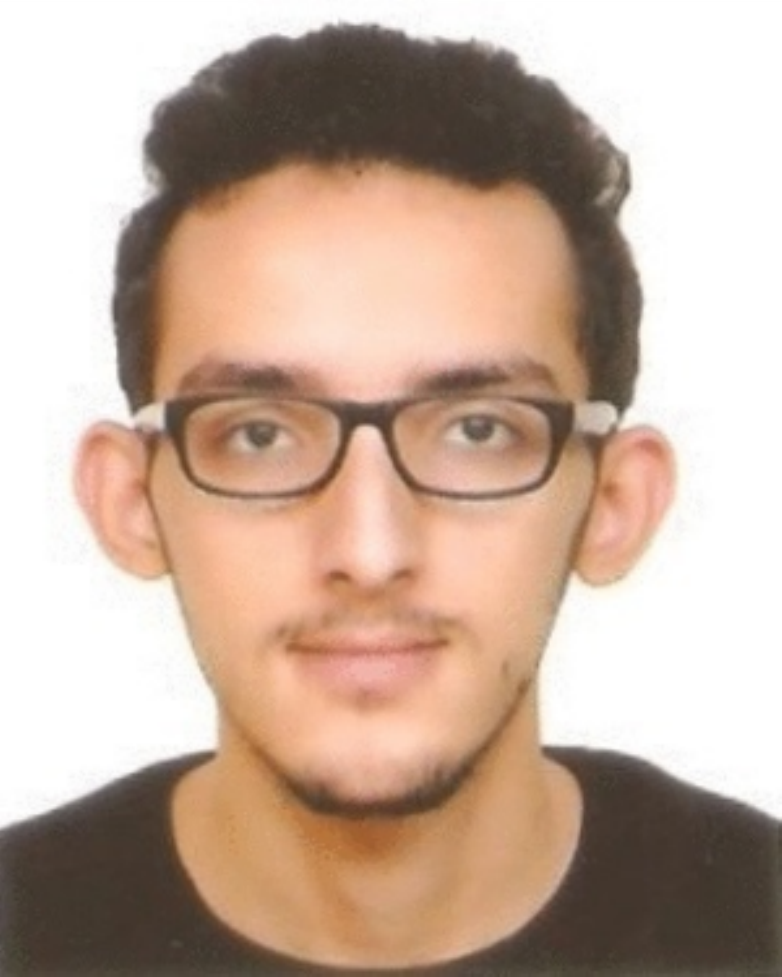}}]{Haythem Chaker}
received the State Engineering degree (Dipl.-Ing.) in Computer Networks and Telecommunications from the National Institute of Applied Sciences and Technology (INSAT), Tunisia in 2019 and the M.Res. in Information Processing and Complexity of Living Systems (TICV) from the National Engineering School of Tunis (ENIT), Tunisia, in co-graduation with the M.Sc. in Mathematics and Computer Science from Paris University, France, in the same year. In 2020, he joined the Signal Processing and Satellite Communications (SIGCOM) research group at SnT, University of Luxembourg as a Ph.D. candidate. His research interests are in wireless systems prototyping with focus on dynamic beamforming design and optimization through the use of new digital signal processing techniques and hardware demonstrations.
\end{IEEEbiography}

\begin{IEEEbiography}[{\includegraphics[width=1in,height=1.25in,clip,keepaspectratio]{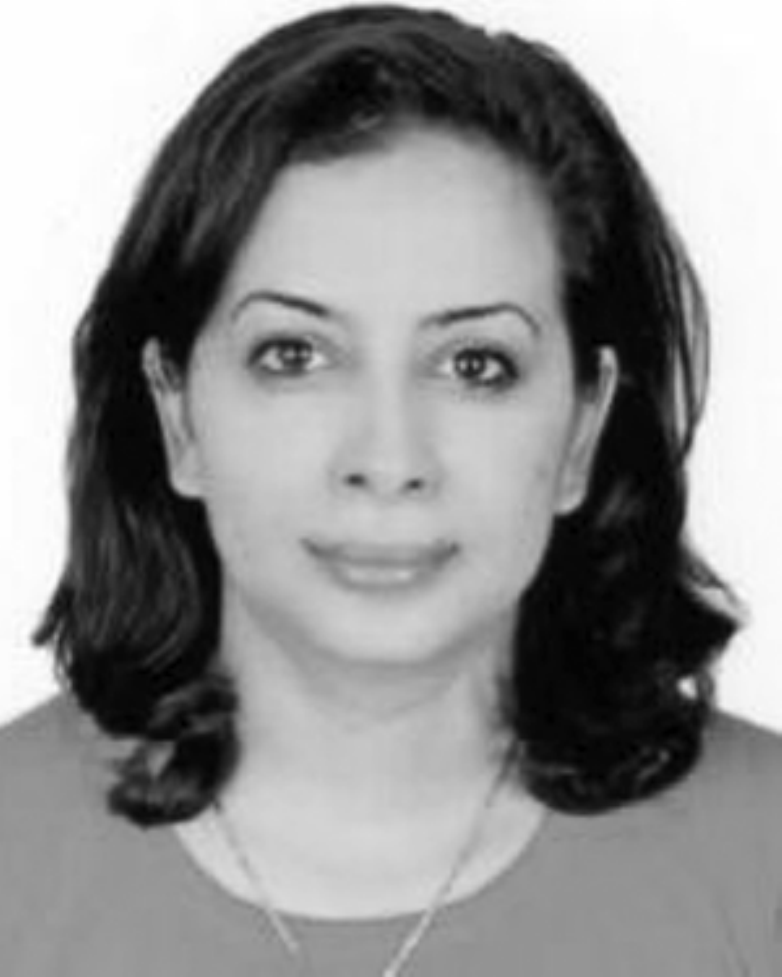}}]{Soumaya Hamouda}
received the Engineering degree in electrical engineering and the DEA (M.Sc.) degree in communication systems from the National Engineering School of Tunis (ENIT), Tunisia, in 1998 and 2000, respectively, and the Ph.D. degree and the “Habilitation Universitaire” degree in technologies of information and communication from the Telecommunications Engineering School of Tunis (Sup’Com), Tunisia, in 2007 and 2015, respectively. She is currently an Associate Professor in telecommunications, a coordinator of the Professional Master in Telecommunication Network Technologies, Faculty of Sciences of Bizerte (Tunisia), the Head of the National Engineering Studies in Telecommunications Committee, and also a member of the Research Laboratory in radio mobile networks and multimedia (MEDIATRON), Sup’Com. Her research interests include radio resource allocation, PHY/MAC protocols and mobility management in wireless networks, 5GNR, mmWaves, IoT in precision agriculture, and e-Health areas. She participated in several national and international research projects with INRIA-Rennes (France), ETRI (South Korea), UAB/CTTC (Spain), and University of Pretoria (South Africa).
\end{IEEEbiography}

\begin{IEEEbiography}[{\includegraphics[width=1in,height=1.25in,clip,keepaspectratio]{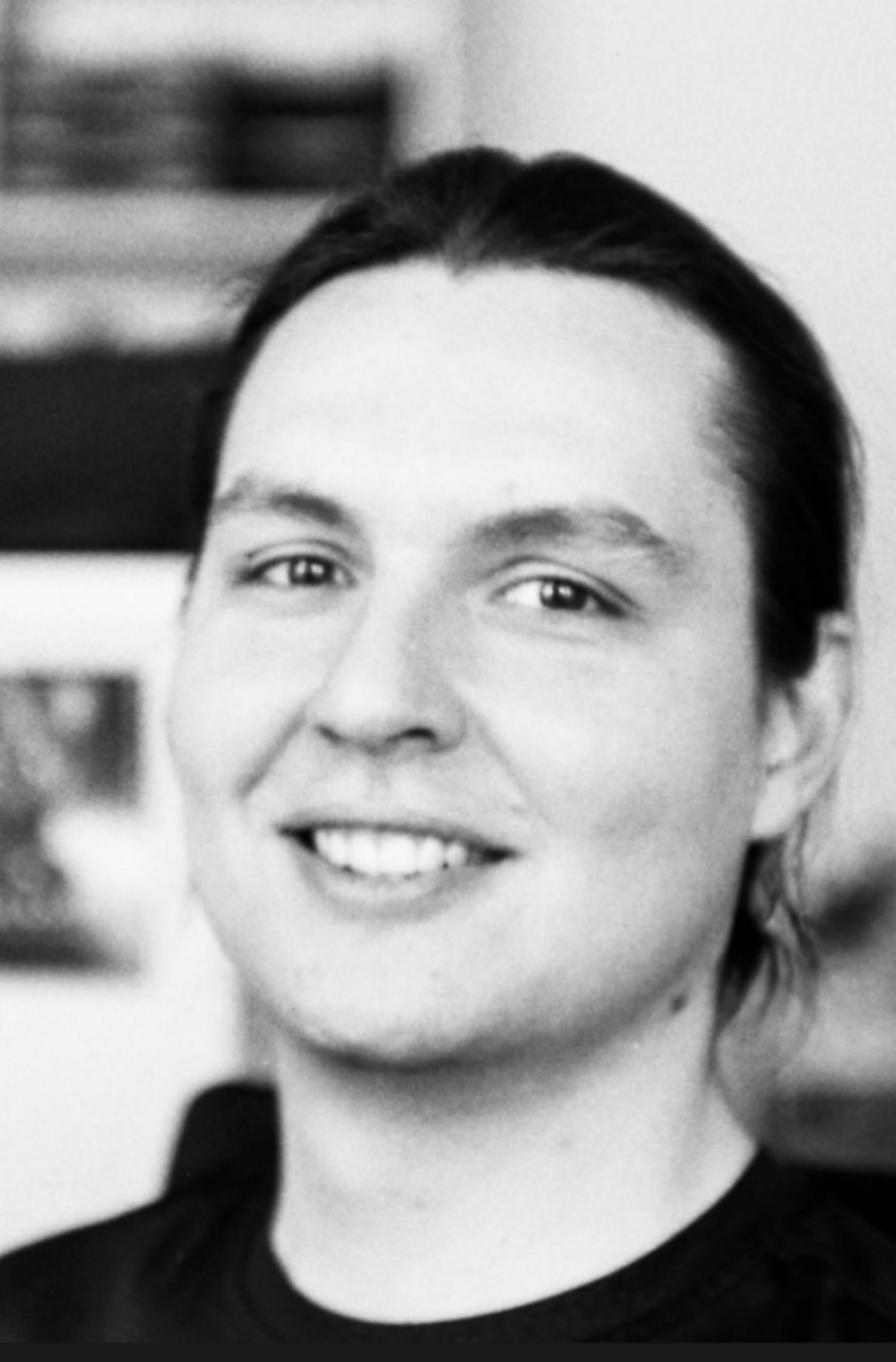}}]{Nicola Michailow}
received the Dr.-Ing. degree in electrical engineering from Technische Universität Dresden in 2015, where he researched non-orthogonal waveforms and flexible numerologies for the 5G PHY. From 2015 to 2018, he was with National Instruments, working on SDR-based prototyping platforms for 4G, 5G and WiFi. Since 2018, he is advancing industrial 5G at Siemens Technology. His current research interests include integrated communications and sensing, 6G and embodied AI systems. He participated in the research projects 5GNOW, CREW, ORCA, IC4F, KICK and Hexa-X.
\end{IEEEbiography}

\vfill

\end{document}